\documentclass[showpacs,preprint,aps,nofootinbib,superscriptaddress]{revtex4}
\usepackage{epsfig}
\usepackage{bm}
\begin{document}

\title{Hadronic parity violation in $\vec{n} p \rightarrow d \gamma$
with effective field theory}

\author{C. H. Hyun}
\email{hch@meson.skku.ac.kr}
\affiliation{Department of Physics and Institute of Basic Science,
Sungkyunkwan University, Suwon 440-746, Korea}

\author{S. Ando}
\email{sando@meson.skku.ac.kr}
\affiliation{Department of Physics and Institute of Basic Science,
Sungkyunkwan University, Suwon 440-746, Korea}

\author{B. Desplanques}
\email{desplanq@lpsc.in2p3.fr}
\affiliation{LPSC, Universit\'{e} Joseph Fourier Grenoble 1,
CNRS/IN2P3, Institut National Polytechnique de Grenoble,
F-38026 Grenoble Cedex, France}

\date{May 31, 2007}

\begin{abstract}

The parity-violating nucleon-nucleon ($NN$) potential
is considered up to next-to-next-to leading order in
heavy-baryon chiral perturbation theory.
We include one-pion exchange at leading order and
the two-pion exchange and two-nucleon contact terms at
next-to-next-to-leading order.
The effects of intermediate (two-pion exchange) and
short-range (two-nucleon contact) terms are probed
by calculating the photon asymmetry $A_\gamma$ in
$\vec{n} p \rightarrow d \gamma$ employing
Siegert's theorem and an accurate phenomenological
potential for the parity-conserving $NN$ interaction.
We explore in detail the uncertainties due to the
parameters that control the contribution of the 
short-range interaction. We obtain about 20\% uncertainty
in the value of $A_\gamma$ up to the next-to-next-to
leading order. We discuss the implication of this uncertainty
for the determination of the weak pion-nucleon coupling constant 
and how the uncertainty can be reduced.

\end{abstract}


\maketitle

\section{Introduction}
\label{intro}
The perturbative calculation of  
the strong nucleon-nucleon ($NN$) potential 
in the framework of effective-field theory (EFT) 
was first suggested by Weinberg \cite{weinberg}.
Counting rules 
systematically arrange
the magnitude of 
a two-nucleon irreducible diagram for
calculating the $NN$ potential 
in terms of 
$Q/\Lambda_\chi$ where $Q$ denotes
a typical small momentum and/or pion mass $m_\pi$
and $\Lambda_\chi$ the chiral scale. 
Accordingly, 
one-pion exchange (OPE) and a constant two-nucleon
contact term are the most dominant contributions.
At the next-to-next-to-leading order (NNLO), there are
two-pion exchange (TPE), contact terms with two
derivatives and/or pion mass factors, 
relativistic corrections, {\it etc}.
If one goes to higher orders,
there appear multi-pion exchanges and
contact terms with more than two derivatives and/or $m_\pi$,
and heavy-meson exchange possibly comes into play.
At low energies where EFT is applicable, 
a physical process is
mostly governed by the long-range interaction
and it is more or less insensitive to what's happening
in the short-range region.
Thus in most cases, heavy mesons are integrated out
from the Lagrangian 
and their short-range interaction is accounted for by contact terms.
A physical observable calculated with the $NN$ potential 
must be independent of a cutoff value that
is introduced in calculating loop diagrams and 
its transformation to coordinate space.
This can be achieved by contact terms together 
with some renormalization method.
Strong EFT potentials thus obtained were applied to the 
analysis of $NN$ scattering phase shifts 
\cite{ork_prc96,egm_npa00,e_05}, showing a well-behaved
convergence and predictive power.
The role of the NNLO potential was intensively explored in
\cite{hpm_plb00} where it was shown to be
important in extending the predictability of the EFT
to higher energies. In addition, it gives a correction
to the leading order (LO) potential which is non-negligible 
even at the energy of a few MeV.

The above approach has recently been extended to the parity-violating  
(PV) nucleon-nucleon ($NN$) potential~\cite{zhu_npa05}. 
Since then, it has been used for the calculation 
of observables \cite{hyun_06, liu_06}, 
especially the PV photon asymmetry $A_\gamma$ in
$\vec{n}p \rightarrow d\gamma$ at threshold, 
where it could be of some relevance.

The PV $NN$ potential is obtained by replacing a
parity-conserving (PC) vertex in the strong $NN$
potential with a PV vertex.
Most of the low-energy PV calculations have relied on 
a one-meson-exchange potential with DDH estimates 
for the PV meson-nucleon coupling constants~\cite{ddh80}.
Theoretical estimations of $A_\gamma$ have been 
extensively worked out with this model (see Refs. 
\cite{kaplan_plb1999,bd_plb2001,hyun_plb2001,rocco_prc2004,hyun_epja05} 
for recent ones).
The most elaborate results with various strong-interaction 
models~\cite{bd_plb2001,hyun_plb2001,rocco_prc2004,hyun_epja05} 
turn out to be basically identical. The asymmetry is do\-mi\-nated
by the PV one-pion-exchange potential (OPEP) and 
the heavy-meson contribution is negligible. 
One can thus discuss whether the measurements of $A_\gamma$
could provide us with an opportunity to determine
the weak pion-nucleon coupling constant $h_\pi^1$.
Some literature has also investigated the PV two-pion-exchange potential
(TPEP) \cite{bd_plb72,pr_plb73,cd_npb74}.
Its contribution to $A_\gamma$, calculated 
with the Hamada-Johnston potential, amounts  
to about $-7$\% of the OPEP contribution \cite{bd_npa75}.
Not surprisingly, the TPEP is also part  
of the pionful EFT approach where it appears at NNLO~\cite{zhu_npa05}.
In the present work, we concentrate on its contribution 
to the asymmetry  $A_\gamma$ mentioned above 
and consider the questions that its estimate raises.

Since the PV asymmetry in $\vec{n}p \rightarrow d\gamma$
is sensitive to the one-pion-exchange contribution,
we adopt 
the heavy-baryon chiral perturbation (HB$\chi$PT).
We employ the Argonne v18 potential for the PC potential 
and the Siegert's theorem for the current operators
and consider the PV potential relevant to
$\vec{n}p \rightarrow d\gamma$ up to NNLO.
This calculation will allow us to estimate the order
and the magnitude of higher-order corrections, which will be
important in pinning down the value of $h^1_\pi$ and its
uncertainty.
At the same time, it will provide a criterion
for the validity of the EFT approach to the PV phenomena.
%
%
\section{Formalism}
\label{sec:formalism}
\begin{figure}
\begin{center}
\resizebox{0.9\textwidth}{!}{%
  \includegraphics{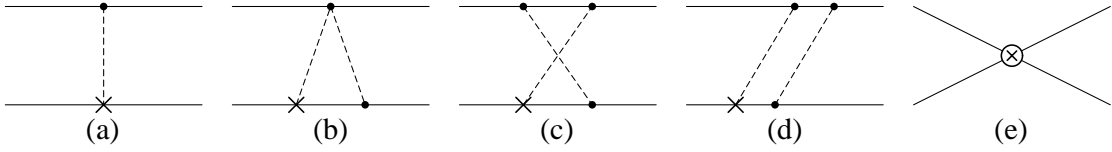}
}
\end{center}
\caption{
Diagrams for PV $NN$ potential up to NNLO: 
(a) for LO (${\cal O}(Q^{-1})$) OPE,
and (b-e) for NNLO (${\cal O}(Q^{1})$) TPE + CT.
Lines (dashed lines) denote nucleons (pions),
vertices with a dot represent PC vertices,
vertices with ``$\bm{\times}$" represent the PV vertex proportional
to $h^1_\pi$, and a vertex with ``$\bm{\otimes}$" denotes
the $NN$ contact vertex function proportional
to the coefficient $C^R_6$.
}
\label{fig:pep}
\end{figure}

%
%
In HB$\chi$PT,
the order of a diagram is counted in terms of a small momentum
$Q$ (more precisely $Q/\Lambda_\chi$) with the following rules;
i) a meson propagator is counted as $Q^{-2}$,
ii) a nucleon propagator as $Q^{-1}$,
iii) a loop integral as $Q^4$,
iv) a vertex as $Q^d$ where $d$ is the number of derivatives
at the given vertex.
With these counting rules, diagrams in Fig.~\ref{fig:pep}
constitute the contributions up to NNLO.
One-pion exchange (a),
and the two-pion exchange (b-d) and the contact (CT) (e) terms 
respectively represent LO and NNLO contributions. 

The PV potential is odd in powers of the momentum transfer in momentum
space or the radial vector in the configuration space, which changes
the orbital angular state by an odd number. In order to keep the
wave function of a fermion system antisymmetric, the PV potential
must satisfy the condition $\Delta(L + S + T) =$ even,
where $L$, $S$ and $T$ denote orbital, spin and isospin states,
respectively. 
The change of $S+T$ in the PV potential should be an odd number. 
For a two-fermion system, the possible combinations of the spin and 
isospin operators in the PV potential must thus be
$(\Delta S,\, \Delta T) = (1,0)$ or (0,1).
Among various possible combinations of spin and isospin operators
that give $\Delta (S+T) = 1$, the term relevant to
our estimation of $A_\gamma$ has the following form in momentum space
\begin{eqnarray}
\tilde{V}_i(\bm{q}) = i (\bm{\tau}_1 \times \bm{\tau}_2)^z
(\bm{\sigma}_1 + \bm{\sigma}_2) \cdot \bm{q} \,\,
\tilde{v}_i (q),
\label{eq:vq}
\end{eqnarray}
with $q \equiv |\bm{q}|$ and $\bm{q} = \bm{p}_1 - \bm{p}'_1$,
where $\bm{p}_1$ ($\bm{p}'_1$) is the momentum of a nucleon 1
in the initial (final) state.
The OPE, TPE and CT terms are obtained as
\begin{eqnarray}
\tilde{v}_{1\pi}(q) &=& - \frac{g_A h^1_\pi}{2 \sqrt{2} f_\pi}
\frac{1}{q^2 + m_\pi^2},
\label{eq:vq_ope} \\
\tilde{v}_{2\pi}(q) &=& \sqrt{2}\pi \frac{h^1_\pi}{\Lambda^3_\chi}
\left\{ g_A \tilde{L}(q)
- g^3_A \left[3\tilde{L}(q) - \tilde{H}(q) \right]
\right\},
\label{eq:vq_tpe} \\
\tilde{v}_{\rm CT} &=& C^R_6,
\label{eq:vq_ct}
\end{eqnarray}
with
\begin{eqnarray}
\tilde{L}(q) &=& \frac{\sqrt{q^2 + 4 m_\pi^2}}{q}
\ln \left( \frac{\sqrt{q^2+4 m_\pi^2} + q}{2 m_\pi} \right),
\label{eq:lq} \\
\tilde{H}(q) &=& \frac{4 m_\pi^2}{q^2 + 4 m_\pi^2} \tilde{L}(q)\, ,
\label{eq:hq}
\end{eqnarray}
where $g_A$ is the axial coupling constant, $f_\pi$
the pion-decay constant and $\Lambda_\chi=4\pi f_\pi$.
The constant $C^R_6$ is the renormalized LEC for a $NN$ contact term
$C_6$ in \cite{zhu_npa05}. It subsumes the
role of all the heavy degrees of freedom integrated out from the theory.
We will discuss how to treat the renormalized LEC in our 
investigation after giving the formulae for the potentials in 
configuration space.

The potential in Eq.~(\ref{eq:vq}) transformed to configuration 
space takes the form
\begin{eqnarray}
V_i(\bm{r}) &=& \int \frac{d^3 \bm{q}}{(2 \pi)^3}
\tilde{V}_i(\bm{q})\, {\rm e}^{-i \bm{q}\cdot \bm{r}} \nonumber \\
&=& i (\bm{\tau}_1 \times \bm{\tau}_2)^z
(\bm{\sigma}_1 + \bm{\sigma}_2) \cdot \left[ \bm{p},\, v_i(r) \right],
\label{eq:Vv}
\end{eqnarray}
where $\bm{p}$ is the conjugate momentum of the relative coordinate
$\bm{r} \equiv \bm{r}_1 - \bm{r}_2$.
For an easier numerical calculation, we rewrite Eqs.~(\ref{eq:lq}, \ref{eq:hq})
in the dispersion-relation form as
\begin{eqnarray}
\tilde{L}(q) &=& - \int^{\infty}_{4m^2_\pi}
\frac{dt'}{2 \sqrt{t'}} \sqrt{t'\! -\! 4m^2_\pi} \left(
\frac{1}{t'\!+\!q^2}\! - \!\frac{1}{t'\! -\! 4m^2_\pi} \right), \\
\tilde{H}(q) &=& \frac{4m^2_\pi}{2} \int^{\infty}_{4m^2_\pi}
\frac{dt'}{\sqrt{t'}}\frac{1}{\sqrt{t'-4m^2_\pi}}
\frac{1}{t'+q^2}\,.
\end{eqnarray}
In calculating the Fourier transformation of Eq.~(\ref{eq:vq}),
in order to obtain a convergent, analytical result,
we introduce monopole form factors of the type
$(\Lambda^2-m^2_{\pi})/(\Lambda^2 + q^2)$ 
in  Eq.~(\ref{eq:vq_ope}) 
and  $\Lambda^2/(\Lambda^2 + q^2)$ 
in Eqs.~(\ref{eq:vq_tpe}, \ref{eq:vq_ct}).
This particular choice for the OPEP preserves the long-range part, 
which, otherwise, could not be easily corrected for 
with contact interactions. 
The roles of the form factor and the cutoff are
(i) to make the numerical calculation easier and more efficient,
and (ii) to cut away the high-momentum region where the dynamics
is essentially unknown and whose detail is irrelevant to low-energy processes. 
We emphasize that the cutoff is arbitrary for a part 
and that final results should not depend on its value.
With the form factor, we rewrite the potential in configuration
space as
\begin{eqnarray}
V^\Lambda_i(\bm{r}) = i (\bm{\tau}_1 \times \bm{\tau}_2)^z
(\bm{\sigma}_1 + \bm{\sigma}_2) \cdot
[ \bm{p},\, v^\Lambda_i(r)]\, ,
\end{eqnarray}
where
\begin{eqnarray}
v^\Lambda_{1\pi}(r) &=& \frac{g_A h^1_\pi}{2 \sqrt{2} f_\pi}
 \frac{1}{4\pi r}
({\rm e}^{-m_\pi r} - {\rm e}^{-\Lambda r})\,, 
\label{eq:vr_ope}\\
v^\Lambda_{2\pi}(r) &=& \sqrt{2}\pi \frac{h^1_\pi}{\Lambda^3_\chi}
\left\{ g_A L^\Lambda (r)
- g^3_A \left[3 L^\Lambda (r) - H^\Lambda (r) \right]
\right\},
\label{eq:vr_tpe} \\
v^\Lambda_{\rm CT}(r) &=& - C^R_6 \Lambda^2 
\frac{{\rm e}^{-\Lambda r}}{4 \pi r}\,,
\label{eq:vr_ct}
\end{eqnarray}
with
\begin{eqnarray}
L^\Lambda (r) &=& \frac{\Lambda^2}{8\pi r}
\int^\infty_{4 m^2_\pi} \frac{dt'}{\sqrt{t'}}\sqrt{t'-4m^2_\pi} 
\left(\frac{{\rm e}^{-\sqrt{t'} r} - {\rm e}^{-\Lambda r}}{\Lambda^2 - t'} 
- \frac{{\rm e}^{-\Lambda r}}{t' - 4m^2_\pi}\right), \\
H^\Lambda (r) &=& -\frac{m^2_\pi \Lambda^2}{2 \pi r}
\int^\infty_{4 m^2_\pi} \frac{dt'}{\sqrt{t'}}\frac{1}{\sqrt{t'\!-\!4m^2_\pi}}
\frac{{\rm e}^{-\sqrt{t'} r}\! -\! {\rm e}^{-\Lambda r}}{\Lambda^2\! -\! t'}\,.
\end{eqnarray}
Notice that in the $\Lambda \rightarrow \infty$ limit, $L^\Lambda (r) $ 
is proportional to the difference of two singular terms,
$r^{-3}$ and $\delta({\vec{r}})$ with an infinite coefficient. 
The 3-dimensional integral over $\vec{r}$ is
however finite and has the sign of the contact term.

The form of $C^R_6$ depends on the regularization scheme.
In Ref.~\cite{zhu_npa05}, all the constant terms obtained from
the dimensional regularization ($d =4 - 2 \epsilon$)
of TPE diagrams are included in $C^R_6$, leading to
\begin{eqnarray}
C^R_6({\rm MX}) = 
C_6 - h^1_\pi \frac{\pi\, g_A}{\sqrt{2}\, \Lambda^3_\chi}
(1 - 3 g^2_A) \left[\frac{1}{\epsilon} - \gamma + \ln(4 \pi)
+ 2 \ln\left(\frac{\mu}{m_\pi}\right) + 2\right],
\end{eqnarray}
where the abbreviation MX stands for `maximal subtraction',
$\gamma = 0.5772$ and $\mu$ is the renormalization scale.
In the minimal subtraction (MN) scheme, the renormalized LEC
satisfies the relation
\begin{equation}
C^R_6({\rm MX}) = C^R_6({\rm MN}) + C_{\rm mn}\, ,
\end{equation}
where
\begin{equation}
C_{\rm mn} = - h^1_\pi
\frac{\sqrt{2}\,\pi\, g_A}{ \Lambda^3_\chi} (1 - 3 g^2_A)
\left[ \ln\left(\frac{\mu}{m_\pi}\right) + 1 \right].
\end{equation}
Either $C^R_6({\rm MX})$ or $C^R_6({\rm MN})$ has to be determined
from a calculation using the underlying theory or from experimental data
with good statistics, neither of which is available at present.
We can only consider possible contributions to $C^R_6$, 
leaving for the future the theoretical 
or experimental determination of the remaining part.

A first contribution is suggested by the one-meson exchange model.
In the heavy-meson limit,
the leading order of the $\rho$-meson propagator 
is a constant and the corresponding term    
can be regarded as part of $C^R_6$. It reads
\begin{equation}
C^R_6(\rho) =  \frac{g_{\rho NN} h^{1'}_\rho}{2 m_N m^2_\rho}\,,
\end{equation}
where $h^{1'}_\rho$ is the PV $\rho NN$ coupling constant 
in the potential. A non-vanishing value is obtained for this coupling 
in a soliton model, $h^{1'}_\rho = -2.2 \times 10^{-7}$ \cite{km_npa89}.
Leaving aside consistency problems between different approaches, 
double counting with the TPEP considered here for instance, 
we notice that it compares to the DDH best value 
of $h^1_\pi$ $(= 4.6 \times 10^{-7})$. The corresponding value
of $C^R_6(\rho)$ is $-1.20 \times 10^{-9}\, \mbox{MeV}^{-3}$
in units of this value of $h^1_\pi$.

A second contribution is provided by the MN scheme 
where $C_{\rm mn}$ is treated independently of $C^R_6({\rm MN})$,
which consequently gives $\mu$ dependence in the result. If we choose
$\mu = m_\rho$, we have $C_{\rm mn} = 4.174\times 10^{-8}\, {\rm MeV}^{-3}$
in units of $h^1_\pi$.
This value is one order larger than $C^R_6(\rho)$, and therefore
the effect of $C_{\rm mn}$ and the dependence on $\mu$ can
be non-negligible. 

A third contribution stems from a deeper examination of the TPEP. 
This one has been obtained by removing from the square-box diagram 
the iterated OPEP, ignoring  the cutoff  introduced 
in actual calculations. 
The correction, which involves the whole strong interaction 
and should contribute to make the results cutoff independent, 
is not easy to calculate. This last property can however 
be restored with a minimal contribution which cancels 
the cutoff-dependent term in Eq. (\ref{eq:vr_ope}). 
This one corresponds to 
\begin{equation}
C^R_6(\pi, \Lambda) = -\frac{g_A h^1_\pi}{2 \sqrt{2} f_\pi \Lambda^2}.
\end{equation}
Its value compares to the $C_{\rm mn}$ one for the smallest values 
of the cutoff $\Lambda$ considered here.

We will explore the above contributions in detail
in the analysis of the  photon asymmetry in $\vec{n} p \rightarrow d \gamma$,
$A_\gamma$. This quantity is defined from the differential cross section
of the process as
\begin{eqnarray}
\frac{d \sigma}{d \Omega} \propto 1 + A_\gamma \cos\theta,
\end{eqnarray}
where $\theta$ is the angle between the neutron polarization and
the out-going photon momentum.
Non-zero $A_\gamma$ values arise from the interference 
of opposite-parity transition amplitudes, {\it e.g.} M1 and E1.
At the thermal energy where the process occurs,
lowest order EM operators may suffice, therefore
we consider the 
E1 operator,
\begin{eqnarray}
\bm{J}_{\rm E1} &=& -i \frac{\omega_\gamma}{4}\,
(\tau^z_1 - \tau^z_2)\, \bm{r} \, ,
\label{eq:op_e1}
\end{eqnarray}
where 
$\omega_\gamma$ is 
the out-going-photon energy.
At the leading order of $h^1_\pi$,
$A_\gamma$ is proportional to
$h^1_\pi$, and we can write $A_\gamma$ as
\begin{equation}
A_\gamma = a_\gamma h^1_\pi \, ,
\label{eq:agamma}
\end{equation}
with
\begin{eqnarray}
a_\gamma =
- 2 \frac{{\rm Re}\left({\cal M}_1 {\cal E}_1^*\right)}{|{\cal M}_1|^2},
\end{eqnarray}
where ${\cal E}_1$ and ${\cal M}_1$ are matrix elements of the E1 and M1
transitions, respectively.
Analytic forms of these amplitudes can be found in \cite{hyun_epja05}.
%
\section{Results and Discussion}
Here we present and discuss numerical results corresponding 
to the various contributions outlined in the previous section. 
The OPE and TPE potentials in configuration space are considered 
as well as individual and total contributions 
to the $A_\gamma$ asymmetry. 

\begin{figure}
\begin{center}
\epsfig{file = 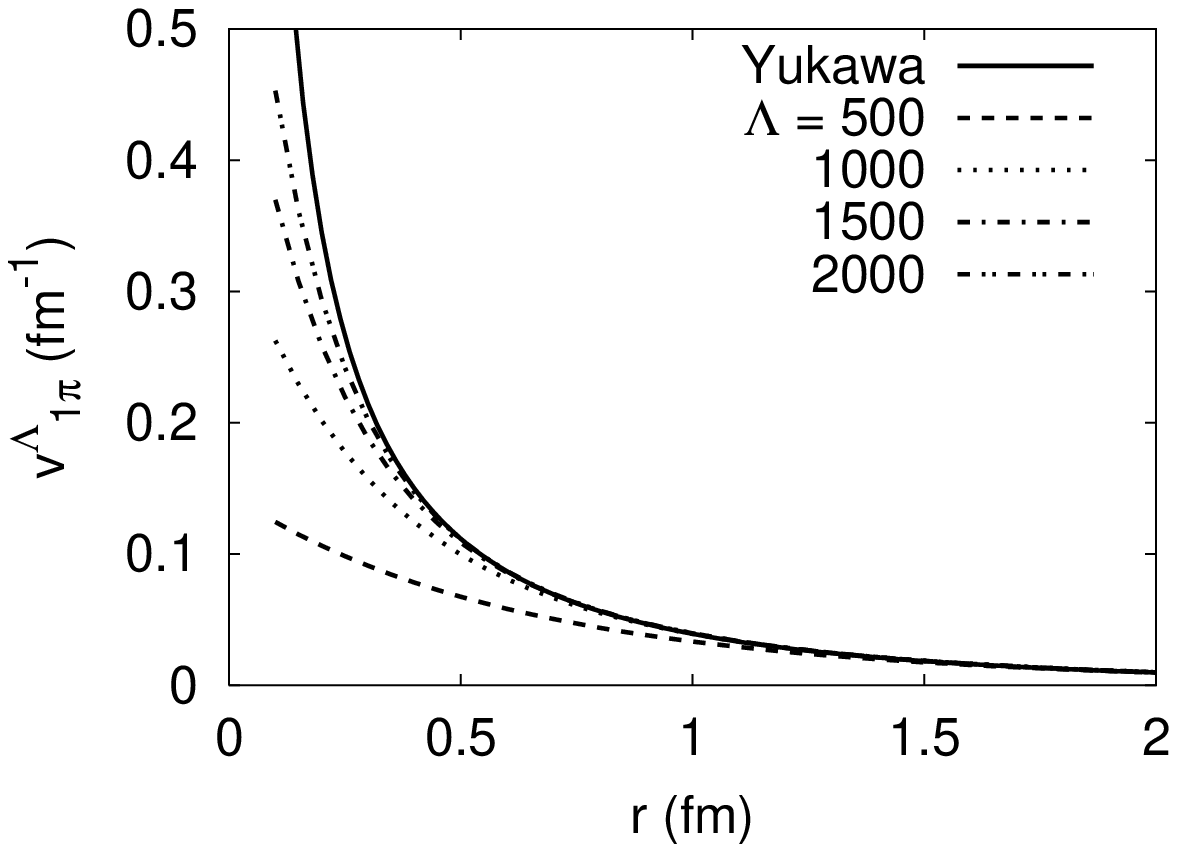, width=7cm}
\epsfig{file = 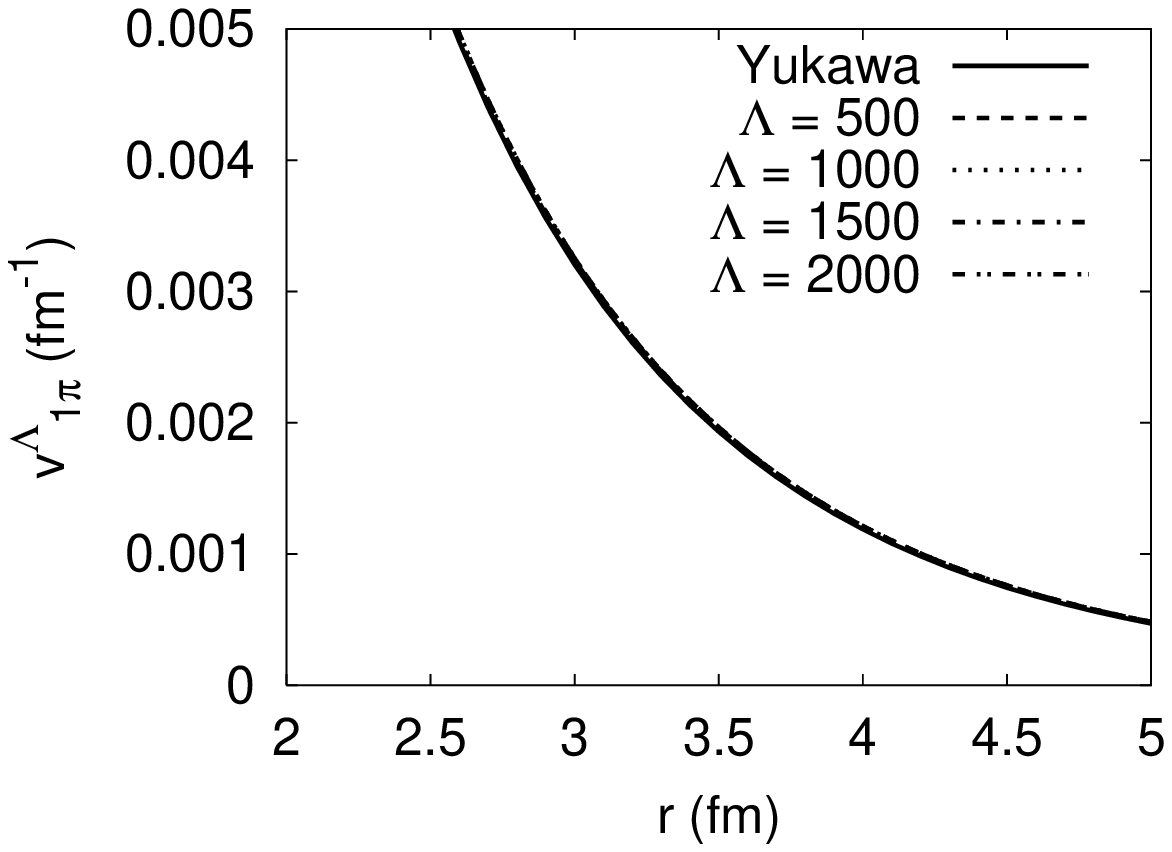, width=7cm} \\
\epsfig{file = 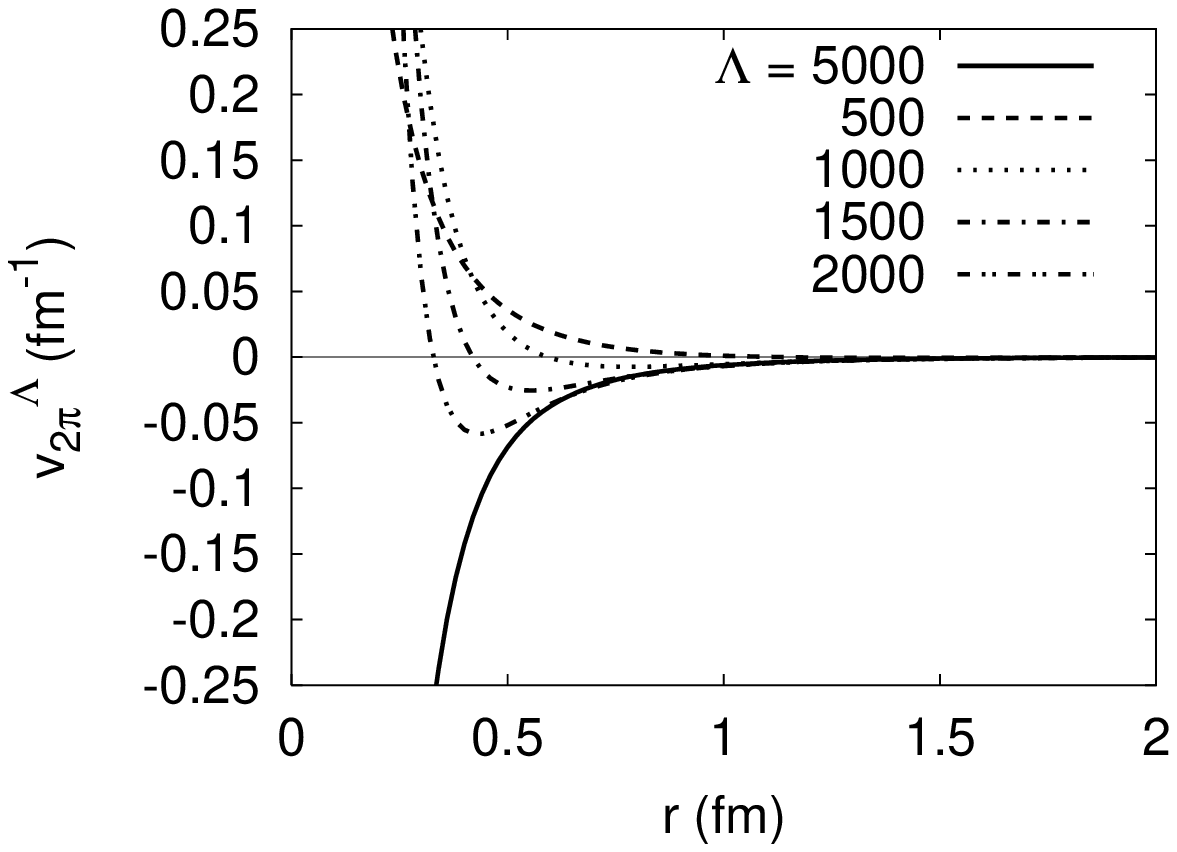, width=7.2cm}
\epsfig{file = 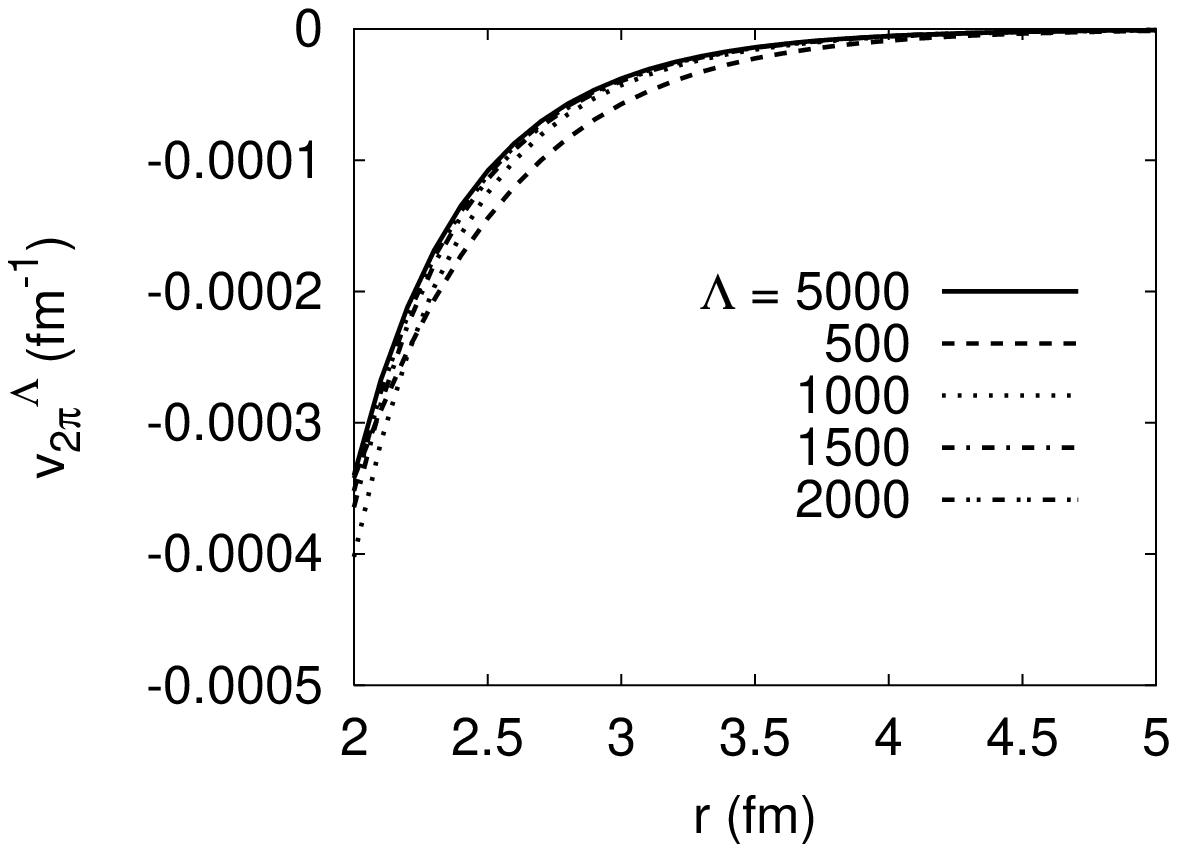, width=7.2cm}
\end{center}
\caption{Potentials $v^\Lambda_{1\pi}(r)$ (upper row) and 
$v^\Lambda_{2\pi}(r)$ (lower row) divided by $h^1_\pi$ in the 
short-intermediate (left column) and long (right column) ranges.
In the figures for TPEP, the result for $\Lambda = 5000$ MeV is shown
as an example of the limiting case $\Lambda \rightarrow \infty$.}
\label{fig:potential}
\end{figure}

The OPEP term $v^\Lambda_{1\pi}(r)$ is shown in the top rows of
Fig.~\ref{fig:potential} in the short (left) and long (right) ranges 
separately.
The curve denoted by `Yukawa' corresponds to the infinite-cutoff value.
Increasing the cutoff $\Lambda$, the potential converges to the Yukawa one. 
While the potential is almost independent of the cutoff value 
in the range $r > 1$ fm,
some sizable deviation from the Yukawa form occurs at $r < 1$ fm 
when the cutoff is equal to 500 MeV.

The TPEP term $v^\Lambda_{2\pi}(r)$ is in the bottom row of
Fig.~\ref{fig:potential}. The long-range behavior is, as OPEP,
independent of the cutoff value, but its magnitude is smaller 
by an order of magnitude. 
Therefore one can conclude that the long-range
dynamics is essentially cutoff independent and governed by OPEP.
On the contrary, TPEP changes sign in the intermediate or
short range (bottom-left panel), and the position
at which the sign change occurs depends on the cutoff value.
What matters in this behavior is that this change occurs 
in the intermediate region where TPEP is supposed to play
an important role. 
The effect of this strong cutoff dependence is investigated
below for the physical observable of interest here.

\begin{table}
\begin{center}
\begin{tabular}{lrrrr}
\hline\noalign{\smallskip}
$\Lambda$ (MeV) & 500 & 1000 & 1500 & 2000 \\
\noalign{\smallskip}\hline\noalign{\smallskip}
$a^{\rm OPE}_\gamma$ & 
$-0.0992$ & $-0.1104$ & $-0.1117$ & $-0.1119$ \\
$a^{\rm TPE}_\gamma$ & 
$-0.0022$ & 0.0073 & 0.0117 & 0.0133 \\
$a^{\rm LEC}_\gamma (\rho)$ &
$-0.0008$ & $-0.0004$ & $-0.0002$ & $-0.0001$ \\
$a^{\rm LEC}_\gamma \mbox{(mn)} $ &
0.0264 & 0.0134 & 0.0065 & 0.0036 \\
$a^{\rm LEC}_\gamma (\pi,\Lambda) $ &
$-0.0128$ & $-0.0016$ & $-0.0004$ & $-0.0001$\\
\noalign{\smallskip}\hline
\end{tabular}
\end{center}
\caption{Contribution to the asymmetry from each term 
as a function of the cutoff $\Lambda$. The renormalization scale
is chosen as $\mu = m_\rho$ in
the minimal subtraction LEC contribution ($\propto C_{\rm mn}$ ).}
\label{tab:agamma_each}
\end{table}

Table~\ref{tab:agamma_each} shows the contribution 
of each term to the coefficient $a_\gamma$ entering 
the expression of the asymmetry $A_\gamma$, Eq. (\ref{eq:agamma}).
The OPEP contribution with $\Lambda = 500$ MeV is slightly smaller 
in magnitude than those with the other cutoff values, but the result 
as a whole is independent of the cutoff value. 
The small dependence of $a^{\rm OPE}_\gamma$ on the cutoff $\Lambda$
can be attributed to the sensitivity of $v^\Lambda_{1\pi}(r)$ 
in the range $r < 1$ fm to $\Lambda$. On the other hand, it seems that
the region below $r \sim 0.5$ fm gives an almost negligible contribution
since $v^\Lambda_{1\pi}(r)$ with $\Lambda =1000$ MeV differs sizably
from its value with larger cutoff values below $r \sim 0.5$ fm, 
but the difference in $a^{\rm OPE}_\gamma$ is negligible.
Roughly speaking, the significant contribution to $a_\gamma$ comes from
the region $r \gtrsim 0.5$ fm.

The TPEP contribution to $a_\gamma$, on the other hand,
varies significantly in sign and magnitude.
The change of $a^{\rm TPE}_\gamma$ can be attributed to the 
strong dependence of $v^\Lambda_{2\pi}(r)$ on the cutoff value
in the intermediate and short ranges. 
The sign change, which can be ascribed to a contact term implied 
by the method used to calculate the potential,  
is physically questionable. 
From the behavior of $v^\Lambda_{1\pi}(r)$ and the sign of
$a^{\rm OPE}_\gamma$, one can deduce that a decreasing term gives
a negative contribution to $a_\gamma$. With a smaller cutoff value,
the decreasing part of TPEP spreads to a larger $r$.
For instance, TPEP becomes increasing at around $r = 0.4$ fm for 
$\Lambda =2000$ MeV, but this occurs around 1.5 fm for $\Lambda = 500$ MeV.
In the discussion of $a^{\rm OPE}_\gamma$, we argued that 
most of the contribution arises from the region 
$r \gtrsim 0.5$ fm. With a smaller cutoff value, the negative contribution
to $a^{\rm TPE}_\gamma$ becomes more substantial and, as a result,
$a^{\rm TPE}_\gamma$ with $\Lambda = 500$ MeV is negative.
The convergence of the TPE contribution with an increasing cutoff value is 
slower than the OPE one, which can also be attributed to the strong
dependence of TPEP on the cutoff value. 
However, $a^{\rm TPE}_\gamma$ will not increase indefinitely and 
we guess that $a^{\rm TPE}_\gamma$ with $\Lambda = 2000$ MeV will
not be much different from that in the limiting case 
$\Lambda \rightarrow \infty$. 
For example, $a^{\rm TPE}_\gamma = 0.0146$ with $\Lambda = 5000$ MeV,
which suggests that the rate of increase with respect to $\Lambda$ 
has decreased compared to that in the smaller $\Lambda$ region.
Thus, without significant error, we can conclude that the TPE contribution
to $a_\gamma$ is about 10\% of the OPE.

All the LEC contributions  
($\propto C^R_6(\rho),\;C_{\rm mn}\;{\rm and}\; C^R_6(\pi, \Lambda) $) 
have a magnitude that decreases with increasing cutoff value. 
With a smaller cutoff value, $v^\Lambda_{\rm CT}(r)$ is less 
short-range peaked and more spread out to larger $r$.
Taking into account the $r \gtrsim 0.5$ fm criterion, 
it is a natural result to have a $a^{\rm LEC}_\gamma$ smaller 
in magnitude with a larger cutoff.
The $C^R_6(\rho)$ contribution is small 
in comparison of the other ones, but this could be a result 
of the particular value of $h^{1'}_\rho$ used 
in its calculation~\cite{km_npa89}. 
The contribution of the $C_{\rm mn}$ term, 
absent in the MX scheme, is comparable to the TPEP one. 
Consequently, the choice of the regularization scheme 
can give a non-trivial effect for the total sum 
of OPE, TPE and LEC contributions,  $a^{\rm tot}_\gamma$. 
The $C^R_6(\pi, \Lambda)$ contribution is also comparable to 
the TPEP one, but only at small $\Lambda$. 
As expected from its derivation, 
it removes the $\Lambda$ dependence of the OPE contribution. 

\begin{table}
\begin{center}
\begin{tabular}{llrrrr}
\hline
$\Lambda$ (MeV) & & 500 & 1000 & 1500 & 2000 \\ 
\hline
$a^{\rm tot}_\gamma \mbox{(MX)}$ & &
$-0.1149$ & $-0.1051$ & $-0.1006$ & $-0.0988$ \\
\hline
&\phantom{a} $\mu = 2 m_\pi$ & $-0.0987$ & $-0.0969$ & $-0.0966$ & $-0.0966$ \\
$a^{\rm tot}_\gamma \mbox{(MN)}$ &\phantom{a} $\mu = m_\rho$ &
$-0.0885$ & $-0.0917$ & $-0.0941$ & $-0.0952$ \\
 &\phantom{a} $\mu = 1$ GeV & $-0.0859$ & $-0.0904$ & $-0.0935$ & $-0.0949$ \\
\noalign{\smallskip}\hline
\end{tabular}
\end{center}
\caption{Total sum of OPE, TPE and LEC contributions in the MX and MN schemes.
Each row of $a^{\rm tot}_\gamma \mbox{(MN)}$ corresponds to $\mu =
2 m_\pi,\, m_\rho$ and 1000 MeV from the top, respectively.}
\label{tab:agamma_total}
\end{table}

In Table~\ref{tab:agamma_total}, we show the sum of OPE, TPE and LEC 
contributions to $a_\gamma$, for the MX and MN regularization schemes. 
Comparison of the results in different
columns gives insight on the cutoff dependence, 
while comparion of rows in the MN case shows how results vary 
with the renormalization scale, $\mu$. 
This scale is conventionally set to the $\rho$-meson
mass, the chiral symmetry breaking scale $\Lambda_\chi$ 
or the order of the cutoff value~$\Lambda$ \cite{e_05}. 
Exploring the corresponding dependence with values 
$\mu = 2 m_\pi,\, m_\rho$ and 1000 MeV,
$a^{\rm tot}_\gamma$ is found to increase with $\mu$ (algebraically).
This is simply related to the contribution of the  $C_{\rm mn}$ term 
which is a logarithmically increasing function of that variable. 
The dependence on the cutoff value is found 
to be opposite in the MX scheme and the MN one with $\mu =  m_\rho$.
Since the contributions of the OPE plus $C^R_6(\pi, \Lambda)$ and  
$C^R_6(\rho)$ terms remain almost constant, 
the $\Lambda$ dependence in the MX scheme is mainly due 
to the TPE one. 
The opposite effect in the MN scheme results 
from the further LEC contribution involving $C_{\rm mn}$, 
which roughly has the same magnitude but is decreasing instead of 
increasing with $\Lambda$. 
The overall $\Lambda$ dependence is sensitive to the
$\mu$ value and can almost vanish ($\mu = 2 m_\pi$) as well 
as be more pronounced ($\mu = $ 1000 MeV).

Because of the considerable dependence on the cutoff $\Lambda$ (MX scheme) 
or on both the cutoff and the renormalization scale (MN scheme), 
the uncertainty of $a^{\rm tot}_\gamma ({\rm MX})$ or
 $a^{\rm tot}_\gamma ({\rm MN})$ is rather large.
Moreover the average value differs from one scheme to the other.
In principle, LEC contributions should make low-energy
observables independent of the input parameters.
In this respect, our results are not quite satisfactory.
In the MX scheme, further contributions are definitively required,  
while in the MN scheme, instead, it appears that an acceptable result 
could be obtained, provided that a low value of $\mu$ is used.
Interestingly, this occurs for a value of $C_{\rm mn}$ that 
roughly compensates the TPEP contribution at short distances, 
which is partly unphysical. A cancellation on this basis 
implies  $\mu=\Lambda$ in the log term contributing to
$C_{\rm mn}$ (and perhaps some modification of the associated constant).
Whatever the approach, results show that OPE is a dominant contribution
to $a_\gamma$, and that the NNLO contributions are in a reasonable
range of about 10 $\sim$ 20 \% of the OPE contributions 
for the most stable cases.
We would say it reasonable in the sense that the range is 
obtained with standard values of input parameters 
and that, at the same time, the amount of the NNLO contribution
has the typical size of higher-order corrections in the
EFT calculation.

\section{Summary}
We have considered the PV $NN$ potential, up to the NNLO,
from heavy-baryon chiral-perturbation theory, 
and applied it to the calculation
of the PV asymmetry in $\vec{n} p \rightarrow d \gamma$
at threshold. 
The OPEP appears at the LO, there is no term at the NLO, and TPEP and CT
terms are picked up at the NNLO.
Heavy degrees of freedom are integrated out by introducing
a monopole form factor in the Fourier transformation of 
the momentum-space potentials and the corresponding contribution 
is ascribed to contact terms.
A renormalized LEC is obtained from the dimensional regularization
and its form depends on the regula\-ri\-zation scheme.
We employ the minimal and maximal subtraction schemes to
determine this form.
Lacking a way to fix the renormalized LEC, we illustrated  
its contribution by two terms.
The first one is obtained from the $\rho$-resonance saturation of the
vector-meson exchange PV potential, and the other one restores the 
$\Lambda$ independence of the OPE contribution. 
In the MN scheme, we have an additional constant term which depends
on the renormalization scale. 
Cutoff dependence is explored with values 
$\Lambda = 500 \sim 2000$ MeV,
and regularization-scheme dependence with renormalization scales
$\mu = 2 m_\pi \sim 1000$ MeV.

The OPE contribution to the asymmetry is satisfactorily cutoff independent.
The TPE potential depends strongly on the cutoff value  
and is somewhat uncertain, in particular in the intermediate range. 
As a result, the TPE contribution to the asymmetry varies widely 
and even shows a change of sign.
However, the TPE contribution is bounded by the limiting cutoff value
$\Lambda \rightarrow \infty$ and does not exceed about 10\% 
of the OPE one.
The LEC contributions are found to be significantly dependent 
on  both the cutoff and renormalization scales introduced 
in their calculations.

The total sum of OPE, TPE and LEC contributions varies non-negligibly,
depending on the choice of the cutoff and renormalization-scale values.
With the parameters considered, we have $a^{\rm tot}_\gamma =
-(0.08 \sim 0.11)$. If the PV $\vec{n} p \rightarrow d \gamma$
experiment measures $A_\gamma$ at the order $10^{-9}$,
the present result of $a^{\rm tot}_\gamma$ can fix the first
digit of the weak $\pi NN$ coupling constant $h^1_\pi$ unambiguously. 
However, some caution is needed.
In principle, the role of LEC is to account for degrees of freedom 
whose microscopic description is  practically irrelevant 
for low-energy processes like the one considered here 
and, at the same time, to  make results 
independent of parameters or regularization schemes employed 
in their calculation.
The present treatment of the LEC
doesn't fully satisfy the latter requirement, 
though some steps in this direction have been considered.
Thus, the estimate of $a^{\rm tot}_\gamma$ made here provides a range
in which its {\it true} value may be, but
it is still premature to claim the result for sure.
For pinning down the {\it true} value of $a^{\rm tot}_\gamma$,
the LEC terms should be treated in a more rigorous and reliable way,
{\it e.g.} fitting them to experimental data if possible, 
obtaining them with lattice calculations, 
or making a better use of renormalization group methods. 
A comparison with earlier TPEP calculations~\cite{bd_plb72,pr_plb73,cd_npb74} 
could also be quite useful. 

\section*{Acknowledgments}
The authors would like to thank S.-L. Zhu, C.~M. Maekawa, U. van Kolck,
M.~J. Ramsey-Musolf and B.~R. Holstein for useful communications about their
published work. 
We are grateful to H. Fearing for reading the manuscript.
We thank the Institute for Nuclear Theory at the University of
Washington for its hospitality and the Department of Energy
for partial support during the completion of this work.
S.A. is supported by Korean Research Foundation and 
The Korean Federation of Science and Technology Societies Grant funded
by Korean Government (MOEHRD, Basic Research Promotion Fund).

\end{document}